\documentclass[aps,prl,showpacs,amssymb,floatfix]{revtex4}
\usepackage{graphicx}

\def\si {\sigma}
\def\de {\delta}
\def\ep {\epsilon}
\def\ef {\epsilon_F}

\def\e2 {\epsilon-\epsilon_k}
\def\be {\begin{equation}}
\def\ee {\end{equation}}
\def\bea {\begin{eqnarray}}
\def\eea {\end{eqnarray}}

\def\om {\omega}

\begin{document}


\title{Theoretical model for the extreme positive magnetoresistance }

\author{George Kastrinakis }

\affiliation{                    
   Institute of Electronic Structure and Laser (IESL), 
Foundation for Research and Technology - Hellas (FORTH), 
P.O. Box 1527, Iraklio, Crete 71110, Greece$^*$ }

\date{9 June 2023}

\begin{abstract} 
We present a model for the positive extreme magnetoresistance (XMR), recently
observed in a plethora of metallic systems, such as PtSn$_4$, PtBi$_2$,
PdCoO$_2$, WTe$_2$, NbSb$_2$, NbP, TaSb$_2$, LaSb, LaBi, ZrSiS and MoTe$_2$.
The model is an extension of our earlier work on positive giant
magnetoresistance, and uses an elaborate diagrammatic formulation.
XMR is a bulk effect (not a surface effect), due
to the dramatic sensitivity of the conductivity to the finite magnetic
field $H$. This is possible at low temperatures, in the presence of finite
disorder elastic spin scattering, and for a special value, predicted
from the theory, of the material-dependent effective Coulomb repulsion.
Good agreement with experiments is obtained. According to our model
XMR is higher in cleaner samples, and anisotropic with regards to the
direction of $H$. We discuss in particular compounds containing the
elements Pt, Sc, and Rh.

\end{abstract}

\pacs{72.10.-d,72.10.Di,72.15.Rn}

\maketitle

{\bf 1. Introduction}
\vspace{0.3cm}

Positive extreme magnetoresistance (XMR) has been observed in several
semimetallic systems like PtSn$_4$ \cite{ptsn}, PdCoO$_2$ \cite{pdco},
WTe$_2$ \cite{wte2,wte2b}, NbSb$_2$ \cite{Nbsb2}, NbP \cite{nbp}, 
TaSb$_2$ \cite{tasb}, LaSb \cite{cavsb,lasb,cavbi,zeng},
LaBi \cite{cavbi,nayak}, MoTe$_2$ \cite{mote2}, 
ZrSiS \cite{zr1,zr2,ZrSiS}, and PtBi$_2$ \cite{ptbi}. Also c.f. the
list of materials displaying positive giant magnetoresistance (GMR) below,
with the distinction between XMR and GMR being rather {\em arbitrary}.
MR is defined as $(\rho(H)-\rho(0))/\rho(0)$, where $\rho(H)$ is the
resistivity which increases in a magnetic field $H$. XMR is unusually
high, reaching even values greater than $10^6$ at low temperatures T of a few
degrees Kelvin. E.g. XMR reaches 1.12 $\times \; 10^7$ for $H=33$ Tesla 
at T=1.8 K in PtBi$_2$ \cite{ptbi}. XMR does not {\em seem} to saturate
with increasing $H$.
Instead, it saturates below a small sample dependent characteristic T,
and decays with increasing T, vanishing below room temperature.
Further, it is {\em highly anisotropic} with respect to the direction of $H$,
hence the Zeeman energy plays a minimum role, if any at all.

Our model in ref. \cite{gmr} explains these features in principle. 
This model successfully explained positive GMR, which 
was observed in several materials, including Pt and Rh \cite{ptf} (already in
1941),
Sc \cite{sc1}, Tl$_2$Ba$_2$CuO$_{6 + \delta}$ \cite{tlba}, 
$\alpha$-(BEDT-TTF)$_2$KHg(SCN)$_4$ \cite{ttf}. After \cite{gmr} was
published, it was also observed in  VO$_x$ \cite{vox}, 
Cd$_3$As$_2$ \cite{cdas}, YSb \cite{ysb}, ScPtBi \cite{scpt},
PtBi$_2$ \cite{ptbi-b},
Cr$_2$NiGa \cite{Cr2NiGa}, and Bi$_2$Te$_3$ \cite{Bi2Te3}.
All the recent XMR experiments bear a striking resemblance to these earlier
GMR experiments. The main difference is the total magnitude of the effect.

Herein we develop further our model, by including additional microscopic
processes via appropriate Feynman diagrams.
In our approach the presence of weak disorder is assumed, necessarily 
including elastic spin scattering from spin-orbit and/or magnetic impurities.

\begin{figure}[tb]
  \includegraphics[width=7truecm]{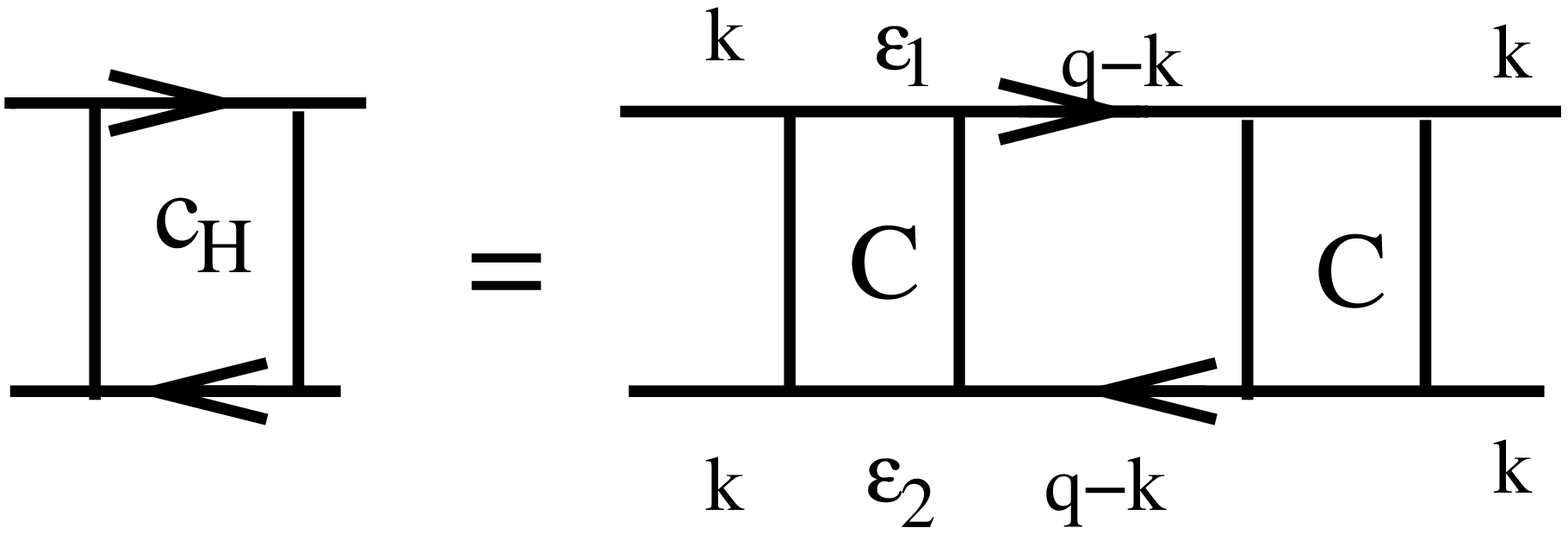}

  \includegraphics[width=11truecm]{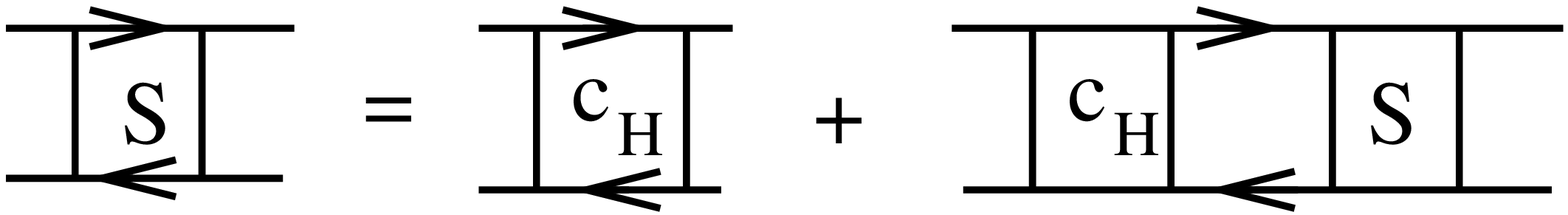}

  \caption{Diagram for $c_H$ given by eq. (\ref{lch}). In our model,
    the dominant dependence on the magnetic field $H$ arises from $c_H$.
    The 2 cooperons are $C_{q,\om}$ with
  $\om=\ep_1-\ep_2$.  Also the ladder diagrams for $S$. In all diagrams,
  the simple horizontal lines stand for the Green's function $G$.}

\label{fig1}
\end{figure}

In a metallic system,
the dimensionless Hubbard interaction constant is $u=U N_F$, where $U$
is the Hubbard constant and $N_F$ the density of states at the Fermi level.
Our mechanism is based on a {\em single} band picture, requiring that 
$u$ be very close to a characteristic value $u_0$ (c.f. eq. (\ref{equ0})
below). Most of the 
materials in which XMR occurs dispose more than one band intersecting 
the Fermi level. Hence it is assumed that $u$ is very close to $u_0$ for
{\em all} these bands.

As pointed out in \cite{cavbi}, XMR is due to a small denominator $\rho(0)$
rather than a big numerator $\rho(H)$. This is consistent with our bulk
mechanism. The conductivity $\si(H)=1/\rho(H)$ is drastically enhanced
for $H=0$. As $H$ increases, $\si(H)$ is quickly reduced, due to the decay
of $c_H$ in eq. (\ref{lch}), thus yielding the effect.

This paper is organized as follows. In section 2 we present the basis
of our model. We introduce the particle-hole propagator $A^j(q,\om)$, which
includes both the Coulomb interaction and elastic spin scattering from the
impurities. In section 3 we develop our basic formalism,
which makes use of diagrams which are symmetric with respect to the
main particle-hole line (which can be seen e.g. in diagrams 2,3).
We also obtain the respective infinite series $W_H$. In section 4 we treat
diagrams which are {\em non-symmetric}, but otherwise very similar to the
ones of section 3. The respective infinite series $X_H$ are also calculated.
In section 5 we calculate the conductivity, based on a resummation of
the series  $X_H$ and $W_H$. A physical interpretation of GMR/XMR as a
manifestation of quantum interference effects at the macroscopic scale is
also offered therein.
In section 6 we consider an even more complete
infinite set of diagrams, and we calculate, and plot,
the respective conductivity.
This section contains also the main discussion on the comparison with
experiments. Section 7 is the Overview.

\vspace{0.6cm} 
{\bf 2. Basis of the model for GMR/XMR} 
\vspace{0.3cm}

The Green's function for the electrons is 
$G^{R,A}(k,\ep)=1/(\ep-\ep_k + \ef \pm i s)$, 
with $s=1/2\tau$. $\ep_k$ is the dispersion relation, $\tau$ the momentum 
relaxation time due to impurities and $\ef$ the Fermi energy (which we take
equal to the chemical potential). In the weak disorder regime 
$\ep_F \tau \gg 1$ \cite{lee}.

In ref. \cite{gmr} we introduced the propagators
$A^j(q,\om)$, $j=\pm 1,0$, in the particle-hole spin-density channel. 
These obey the coupled Bethe-Salpeter equations
$A^1=U+U \; {\cal D}^1 A^1+U\; {\cal D}^0 A^0$ and
$A^0=U \; {\cal D}^0 A^1+U\; {\cal D}^{-1} A^0$. The components of the density 
and spin-density correlation functions are ${\cal D}^{\pm 1} ={\cal D}^{1,\pm 1}$,
${\cal D}^0 = \left\{{\cal D}^{0,0}-{\cal D}^{1,0} \right\}/2$ with 
${\cal D}^{j,m}(q,\om) = N_F \left\{ D q^2+j 4 \tau_S^{-1}/3 \right\}
/ \left\{D q^2+j 4 \tau_S^{-1}/3 -i \om -i m \; \om_H \right\}$. Here
$q$ and $\om$ are the momentum and energy difference between particle and
hole lines,
$D$ the diffusion coefficient, $\om_H$ the Zeeman energy and $\tau_S^{-1}$
the total elastic spin scattering rate due to impurities ($\hbar=c=1$). 
$\tau^{-1}=\tau_0^{-1}+\tau_S^{-1}$, where $\tau_0^{-1}$ is the impurity 
scattering rate without spin-flip. (Also $A^{-1}(\om_H)=A^{1}(-\om_H)$.) 

\begin{figure}[tb]
  \includegraphics[width=8truecm]{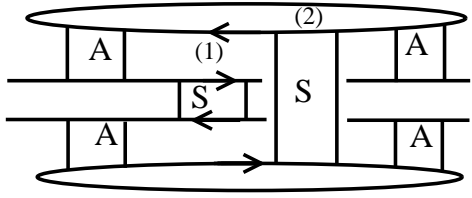}
  \caption{Diagrams including the block $J$ between pairs of propagators
    $A^j$ - c.f. eq. (\ref{eqa}) - in the upper and lower part of the
    diagram. We actually show the 3 different variants of these
diagrams. $J_1 $ corresponds to the case where propagator $S$ marked (2)
is absent, $J_2 $ corresponds to the case where propagator $S$ marked (1) 
is absent, and $J_3 $ corresponds to both $S$'s being present.
The $A^j$'s are {\em not} part of the contribution $J$, as they are
considered separately in eq. (\ref{eqza}) for $Z_{2j}$ etc. These diagrams
are part of the infinite series $W_H$ in eq. (\ref{exwh}).}

\label{fig2}
\end{figure}

The solution of these equations is
\be
A^j(q,\om)=\frac{ K_{uj} \; Dq^2 - i L_{uj} \; \om +M_{uj} }
{ A_u \; Dq^2 - i B_u \; \om + C_u } \;\;   ,
\ee
where in the "dynamic limit"
$D q^2 < \om$ we have \cite{gmr,mit}
\be
A_u=12-20 u + 15 u^2/2 + 6 \Omega_H^2 \;\;, \;\; B_u=4-6u+3u^2/2+2 \Omega_H^2
\;\; , \;\; C_u = r\left\{1-2u+3u^2/4 + \Omega_H^2 (1-u^2/4) \right\}  \;\; .
\label{eqa}
\ee
Here $r=4 \tau_S^{-1}/3$ and $\Omega_H=3 \om_H \tau_S/4$.
$M_{u0}=U u r(1+\Omega_H^2)$, $M_{u1}=U r (1-u)$. In the following, we will
assume that the dependence on $\om_H$ is {\em negligible}. For $\om_H=0$
we have
\be
B_u=0 \text{ for } u=u_0=0.845 \;\; .  \label{equ0}
\ee
We show below that this gives rise to an enhancement (``resonance'') factor
proportional to $1/B_u$, c.f. eq. (\ref{reso}), which 
yields finally the extra-ordinary positive 
extreme magnetoresistance \cite{gmr}. For metallic systems it is often 
appropriate to consider the effective RPA-corrected
value $u_{eff}=u/(1-u^2)$, which is to be compared to $u_0$ \cite{gmr}.
This $u_{eff}$ was found in \cite{gmr} to agree with $u_0$ for Sc, Pt and
Rh, as calculated from first principles \cite{siga}. We also note that {\em 
no such enhancement} arises in the "static limit" $D q^2 > \om$.

$ C(q,\om)=1/\{ (2\pi \; N_F \; \tau^2)(Dq^2-i \om) \}$ is the
cooperon propagator in the particle-particle channel \cite{lee,kawa}. 
We consider the relevant contribution
$c_H = \sum_q \; C^2(q,\om=0)$ - c.f. fig. 1. In $d=3$ dimensions we have
\be
c_H = \frac{1}{(2\pi)^3 N_F^2 \tau^4  \sqrt{D_i}}    \label{lch}
\sum_{n=0}^{\infty} \left\{ \frac{ \sqrt{s}}{ a_n^2} \frac{1}{a_n^2+s}
+\frac{1}{a_n^3} \tan^{-1}\left(\frac{\sqrt{s}}{a_n} \right) \right\}  \;\;,\;\;
\ee
where $a_n^2=4 D_{i \perp} e H (n+1/2) + \tau_{\phi}^{-1}(T)$
\cite{kawa,gmr}. Here $H$ is along the axis $i$ and $i \perp$ stands
for the plane perpendicular 
to the axis $i$. In our model, the dominant
dependence on the magnetic field $H$ arises from $c_H$.
$\tau_{\phi}^{-1}(T)=\alpha + \beta \; T^p$ $(p \geq 1)$
is the electron dephasing rate, which
increases fast with $T$, thus leading to the decay of the cooperon, and
yielding a {\em vanishing MR} (due to this mechanism) at higher $T$.
We have shown \cite{gkdef} that $\alpha > (2/3)\tau_{sp}^{-1}$, where
$\tau_{sp}^{-1}$ is the elastic scattering rate due to {\em magnetic}
impurities only.
We also overall assume $\tau_S^{-1} < \tau_{\phi}^{-1}(T)$.
Note that external lines have momentum
$k$, while in between the 2 cooperons they have $q-k$, where 
$q<1/\sqrt{D \tau} \ll k_F$, with $k_F$ the Fermi momentum. Note that 
the most relevant momentum range is $k \sim k_F$ for the Green's functions.  
Here $\Gamma(k,\ep_1,\ep_2) = G(k,\ep_1) G(k,\ep_2)$. 
We consider the respective infinite ladder series of $c_H$ - c.f. fig. 1
\be 
S(k,\ep_1,\ep_2) =   c_H \; \Gamma(k,\ep_1,\ep_2) \; 
\sum_{n=0}^{\infty} \; \left( c_H \; \Gamma^2(k,\ep_1,\ep_2) \right)^n
=\frac{ c_H \; \Gamma(k,\ep_1,\ep_2) }
{ 1 - c_H \; \Gamma^2(k,\ep_1,\ep_2) } \;\;.\;\;  \label{eqs}
\ee

\begin{figure}[tb]
  \includegraphics[width=8truecm]{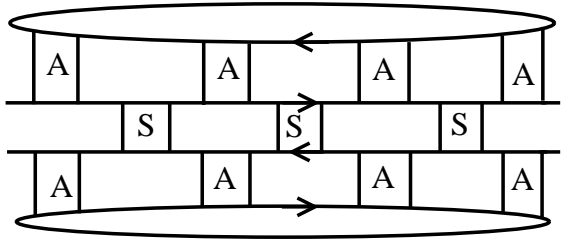}
  \caption{Higher order diagram with $2\times 2n=8$ $A^j$'s, i.e. $n=2$.
    To simplify the presentation, {\em only one propagator $S$}
    is shown between the $A^j$'s. In reality it is possible to have one or
    two $S$'s between the $A^j$'s, as in figure 2. All similar diagrams
    with up to $n=\infty$ are considered and summed up. These diagrams
are part of the infinite series $W_H$ in eq. (\ref{exwh}). }

\label{fig3}
\end{figure}

Previously \cite{gmr} we only considered the first term of this series.
We have 3 contributions, from the respective diagrams shown in fig. 2.
\bea
J_1(k,\ep_1,\ep_2,\om_1,\om_2)= S(k,\ep_1,\ep_2) \; \Gamma^2(k,\ep_1,\ep_2)
 \; \Gamma(k,\ep_2+\om_2,\ep_1-\om_1)  \;\;,  \\
J_2(k,\ep_1,\ep_2,\om_1,\om_2)= S(k,\ep_2+\om_2,\ep_1-\om_1) \;  
\Gamma^2(k,\ep_2+\om_2,\ep_1-\om_1) \; \Gamma(k,\ep_1,\ep_2)   \;\;,  \\
J_3(k,\ep_1,\ep_2,\om_1,\om_2)= S(k,\ep_1,\ep_2) \; S(k,\ep_2+\om_2,\ep_1-\om_1) 
\; \Gamma^2(k,\ep_1,\ep_2) \; \Gamma^2(k,\ep_2+\om_2,\ep_1-\om_1)  \;\;.
\eea
The total contribution equals
$J(k,\ep_1,\ep_2,\om_1,\om_2)=J_1(k,\ep_1,\ep_2,\om_1,\om_2)
+J_2(k,\ep_1,\ep_2,\om_1,\om_2)+J_3(k,\ep_1,\ep_2,\om_1,\om_2)$.

In sections 3 and 4 we account systematically for relevant diagrams
containing the particle-hole propagator $A^j(q,\om)$ and
$J(k,\ep_1,\ep_2,\om_1,\om_2)$.

\vspace{0.6cm}
{\bf 3. Symmetric in $A^j(q,\om)$ diagrams}
\vspace{0.3cm}

We consider appropriate diagrams, as in fig. 2. We show the calculation for 2 
$A^j(q,\om)$'s on the upper part of the diagram, and another 2 in the lower
part. First, we will {\em only} consider diagrams with {\em mirror
  symmetric upper and lower parts}. Afterwards we will also
consider diagrams which are {\em not mirror symmetric} with regards to the 
central particle-hole lines. The latter will
turn out to be {\em sub-dominant}.  Explicitly for fig. 2

\bea
w_{2j} = T^4 \; \sum_{q_1,\om_1} \left( A^j(q_1,i \om_1) \right )^2
\; \sum_{q_2,\om_2} \left( A^j(q_2,i \om_2) \right )^2 
\; \sum_{k,\ep_1,\ep_2} \;
J(k,i\ep_1,i\ep_2,i\om_1,i\om_2) \;
\;\; . \;\; 
\eea
Here $\om_1, \om_2$ and $\ep_1,\ep_2$ are Matsubara energies for
bosons and fermions, respectively, and 
$\ep_1 (\ep_1-\om_1)<0$, $\ep_2(\ep_2+\om_2)<0$.

We first consider
\bea
F_2(i\om_1)=T\sum_{\ep_1} \; J(k,i\ep_1,i\ep_2,i\om_1,i\om_2)
=\frac{1}{2 \pi i} \int_{-\infty}^{+\infty} d\ep \; 
\left( n_F(\ep) - n_F(\ep-i\om_1) \right)  \; 
J(k,\ep,i\ep_2,i\om_1,i\om_2) 
\;\;,
\eea
with $n_F(x)=(1/2)$tanh$(x/2T)$ \cite{agd}.
Then we carry out the summation
\bea
Z_{2j} =T \;\sum_{q_1,\om_1} \;\left( A^j(q_1,i \om_1) \right )^2 \; F_2(i\om_1)
= \frac{1}{2 \pi i} \int_{-\infty}^{+\infty} d\om \; n_B(\om)\;
\sum_{q_1} \left( A^j(q_1,\om) \right )^2 \; F_2(\om) 
\simeq \frac{s_j}{2 \pi i} \int_{\om_a}^{\om_b} d\om \;   \label{eqza}
\frac{ n_B(\om)\;  F_2(\om)} {B_u \; \om}  \;\;, \;\; 
\eea
where $n_B(x)=(1/2)$coth$(x/2T)$ \cite{agd}, 
$s_j= M_{uj}^2 \sqrt{|C_u|}/\{4\pi^2 (A_u D)^{3/2}\}$ for $d=3$,
and $\om_a<\om_b$ are both of the order of a fraction of $\ef$.
Use was made of the equality
\be
z_{2n} = \int_{x_0-\de}^{x_0+\de} \frac{f(x) \; dx}{(x-x_0-i \de)^{2n}} \simeq 
\frac{g_n \; f(x_0)}{(2n-1) \; 2^{n-1} \; \de^{2n-1} }
\;\; ,  \label{reso}
\ee
with $|g_n|=1$ \cite{gm1},
$\de = B_u \; \om \rightarrow 0$ for $u \rightarrow u_0$, 
$x = A_u D q^2$, and $x_0=-C_u>0$.
The summations over $q_2,\om_2$ proceed in the same way.

Finally, with $\om_0=\om_b-\om_a$ and considering 
$ n_F(\ep) - n_F(\ep-\om) \simeq \om \; \partial n_F(\ep)/\partial \ep$
($T \rightarrow 0$) we obtain 
\be
Z_{2j}=\frac{ s_j \om_0}{4\pi^2 \; B_u} J(k,\ep=0^+,i\ep_2,0,i\om_2) \;\;.
\ee
Carrying out the summations over $q_2,\om_2$ and $\ep_2$ yields
\be
w_{2j}=\left( \frac{ s_j \om_0}{4\pi^2 \; B_u} \right)^2 \; 
\; \sum_k \; J(k,\ep=0^+,\ep=0^-,0,0) \;\;.
\ee

We want to evaluate 
$P = \sum_{k<k_F} \; J(k,0^+,0^-,0,0) = 2 h_a + h_b$,
with the contributions $2 h_a$ (due to 
$J_1+J_2$) and $h_b$ (due to $J_3$).
Now we take $S_k=S(k,\ep=0^+,\ep=0^-)$ and
$\Gamma_k=\Gamma(k,\ep=0^+,\ep=0^-)$.
With
\be
X_k = (\ep_k-\ef)^2+s^2  \;\; , \;\;
\ee
we have
\be
h_a = \sum_{k<k_F} S_k \; \Gamma^3_k =
\sum_{k<k_F} \frac{c_H}{X_k^2-c_H}\; \frac{1}{X_k^2}
\;\; , \;\; 
h_b = \sum_{k<k_F} S^2_k \; \Gamma^4_k =
\sum_{k<k_F} \frac{c_H^2}{\left(X_k^2-c_H\right)^2}\; \frac{1}{X_k^2}
\;\; . \;\; 
\ee
We take $\sum_{k<k_F} = N_F \; \int_0^{\ef} \; d\ep_k$. Then
the two integrals can be calculated exactly using a symbolic manipulation
package. We use Mathematica throughout. With 
$r_{\pm}=\sqrt{s^2 \pm \sqrt{c_H}}$, the total result is
\bea
P = N_F \left\{ \frac{2}{\sqrt{c_H}} 
\left[\frac{1}{r_-}\tan^{-1}\left(\frac{\ef}{r_-}\right)
- \frac{1}{r_+}\tan^{-1}\left(\frac{\ef}{r_+}\right) \right]
+\frac{7 \sqrt{c_H} + 6 s^2}{8 r_+^3 \; \sqrt{c_H} } \; 
\tan^{-1}\left( \frac{\ef}{r_+} \right)  \right. \label{eqp} \\  \left.
+\frac{7 \sqrt{c_H} - 6 s^2}{8 r_-^3 \; \sqrt{c_H} } \; 
\tan^{-1}\left( \frac{\ef}{r_-} \right)
+\frac{1}{4} \frac{\ef}{c_H-s^4} 
\frac{c_H+s^2(s^2+\ef^2)}{c_H-(s^2+\ef^2)^2}
-\frac{\ef}{2 s^2} \; \frac{1}{s^2+\ef^2}
-\frac{1}{2 s^3} \; \tan^{-1}\left( \frac{\ef}{s} \right) \right\}
\;\;. \;\;   \nonumber
\eea

\begin{figure}[tb]
  \includegraphics[width=8truecm]{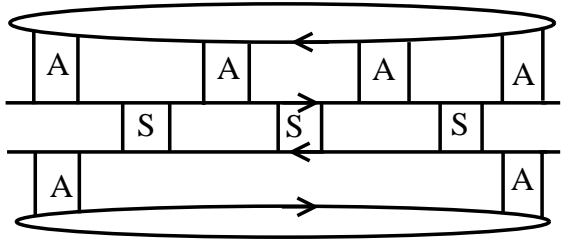}
  \caption{ Non-symmetric diagram. Here two $A^j$'s are missing
    below the main particle-hole lines. The outermost $A^j$'s
    {\em are always kept} in such diagrams. As in fig. 3, to simplify
    the presentation, {\em only one propagator $S$}
    is shown between the $A^j$'s. In reality it is possible to have one or
    two $S$'s between the $A^j$'s, as in figure 2. This diagram
is part of the infinite series $X_2(z)$ in eq. (\ref{exxh}).  }
    
\label{fig4}
\end{figure}

For higher orders $n>1$, as in fig. 3, it is easy to show that the
respective result is 
$w_{2n,j}= r_0 \; M_{uj}^{4n} \; P^{2n-1}/\left\{ 2^{2n -2} \; (2n-1)^2
\; B_u^{4n-2} \right\}$,
with $r_0=\om_0^2 \; |C_u| / \left\{ 16\pi^4 (A_u D)^3 \right\} $.

Now, the ratio $\Lambda_u=M_{u0}/M_{u1}=\sqrt{s_0/s_1}$ equals 5.45 and 4.07 
for $u=u_0$ and 0.95 $u_0$, respectively. 
It is easy to consider all diagrams which contain factors of the type
$\left \{ \left( A^0(q,\om) \right)^4 + \left( A^1(q,\om) \right)^4  
\right \}^n$ for any $n \geq 1$. For this purpose we consider the factor
(in which the numerator of $A^j(q,\om)$ is approximated by $M_{uj}$)
\be
l_0^4 = M_{u0}^4 \; \left( 1+1/\Lambda_u^4 \right)  \;\;.
\ee
Considering $\Lambda_u = 5$,  $1/\Lambda_u^4 = 1/625$ is a small
number.
Then, with $w_{2n}$ containing the contributions from all these diagrams,
we have the series \cite{seri1}
\bea
W_{1H} = \sum_{n=1}^{\infty} w_{2n} 
=2 r_0 \; \sum_{n=1}^{\infty} \frac{P^{2n-1} \; l_0^{4n}}{2^{2 n} \; (2n-1)^2 \;
  B_u^{4n-2}} = r_0\;l_0^2 \; z \; I_1(z) \;\;
\nonumber \\
  I_1(z) = \sum_{m=0}^{\infty} \frac{z^{2m}}{(m+1/2)^2}\;\;, \;\;
z = P \; l_0^2 /(2 \; B_u^2) \;\;. \;\;   \label{w1h}
\eea

We also consider the rest of the diagrams of order $1/\Lambda_u^4$. 
Because there are $4 \times 3^{n-2}$ such terms for
$n>1$ (with $2n-1=3,5,7,$... etc. blocks $P$ from
eq. (\ref{eqp})), we obtain the series
\be
W_{2H} = 8 r_0\;l_0^2 \; z^3 \; I_2(z) \;\;, \;\;
I_2(z) = \sum_{n=0}^{\infty} \frac{3^n \; z^{2n}}{(n+3/2)^2}   \;\;.
\ee
The total contribution due to $W_{1H}$ and $W_{2H}$ is
\be
W_{H} = W_{1H}+W_{2H}/\Lambda_u^4 + O(1/\Lambda_u^8)  \;\; . \label{exwh}
\ee
We also write this as
\be
W_{H} = r_0\;l_0^2 \; z \; I(z) \;\; , \;\;
I(z) = I_1(z) + 8 z^2 \; I_2(z)/\Lambda_u^4 
\;\;. \;\;    \label{finalwh}
\ee

\vspace{0.6cm}
{\bf 4. Non-symmetric in $A^j(q,\om)$ diagrams}
\vspace{0.3cm}

Now we turn our attention to the non-symmetric diagrams (with regards to the 
central particle-hole lines). These are
of the same type as the ones considered above, but with a distinct
difference. There are {\em two $ A^j(q, \om)$'s missing} either in the
upper or in the lower part of the diagrams as in figs. 4 and 5.
In particular, we consider diagrams such that the leftmost and rightmost
$A^j(q, \om)$'s  {\em are kept in place}. The two missing $A^j(q, \om)$'s
would be located anywhere {\em in between} these two external ones.
Also, due to this deletion, two energy summations are removed as well,
which yields an overall factor $i^2=-1$ for these non-symmetric diagrams
- c.f. $X_1(z)$ and $X_2(z)$ below.

The calculation here proceeds in a very similar manner as for the
symmetric diagrams. Of course care has to be taken with all the factors
which appear in an upper-lower non-symmetric way. E.g. all the diagrams
here contain the asymmetry factor $h_m= g_m \; g_{m-1} = (-1)^{m+1}$ 
\cite{gm1}.
The diagrams herein contain the factors $Q_1$ and $Q_2$
shown in figs. 6 and 7 respectively. We have
\be
Q_1 = \sum_{k<k_F} \left\{ 4 \; S^2_k \; \Gamma^5_k   \label{eqq1}
+4 \; S^3_k \; \Gamma^6_k + S^4_k \; \Gamma^7_k \right\} \;\;,
\ee
and
\be
Q_2 = \sum_{k<k_F} \left\{ 8 \; S^3_k \; \Gamma^7_k
+ 12\; S^4_k \; \Gamma^8_k + 6\; S^5_k \; \Gamma^9_k
+ S^6_k \; \Gamma^{10}_k \right\}  \;\;.     \label{eqq2}
\ee
Assuming a constant density of states $N_F$, these integrals can be
calculated exactly, as we did for $P$ above.

\begin{figure}[tb]
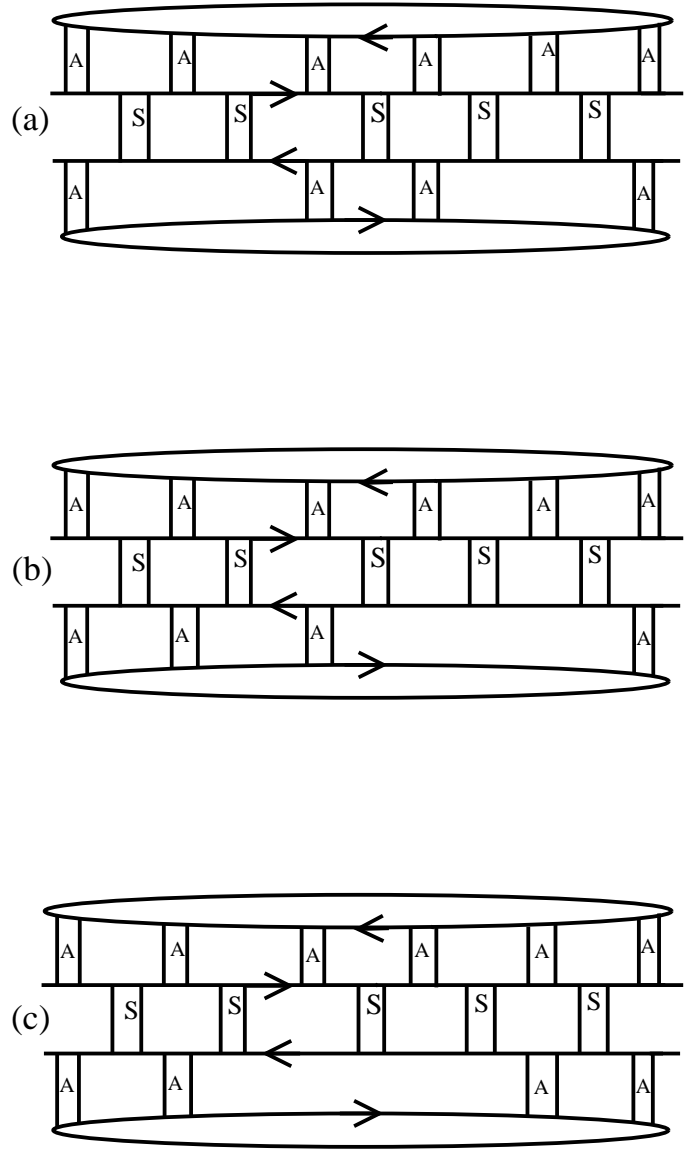

  \includegraphics[width=9truecm]{xfig5a.eps}
  \includegraphics[width=9truecm]{xfig5b.eps}
  \includegraphics[width=9truecm]{xfig5c.eps}

  \caption{ Non-symmetric diagrams. Two $A^j$'s are missing
 below the main particle-hole lines. Note that fig. (a), (b),
 and (c) differ in the positions of missing $A^j$'s.
 All these diagrams with $2 m +2(m-1)=4m-2$ $A^j$'s are taken into account
 with up to $m=\infty$. Diagrams (a) are part of the infinite series $X_1(z)$
 in eq. (\ref{exxh}), while diagrams (b) and (c) are part of the infinite
 series $X_2(z)$ in eq. (\ref{exxh}).  }
    
\label{fig5}
\end{figure}

\begin{figure}[tb]
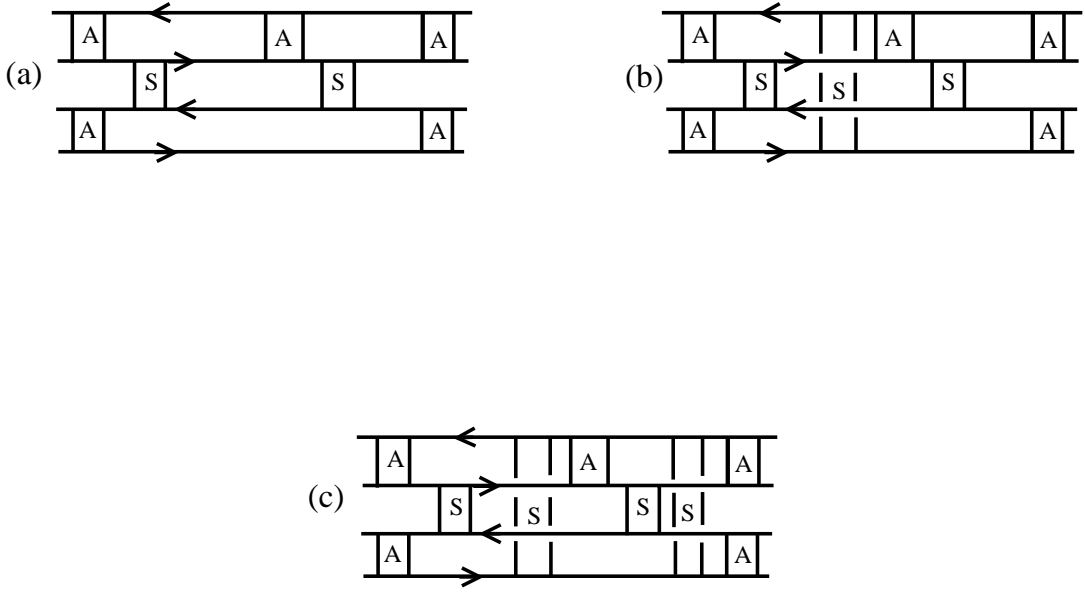

  \includegraphics[width=8truecm]{xfigq1a.eps}
  \includegraphics[width=8truecm]{xfigq1b.eps}
  \includegraphics[width=8truecm]{xfigq1c.eps}
  
  \caption{ Diagrams for the {\em internal} part $Q_1$, i.e. without
    the outer bubble lines. Three distinct contributions (a)-(c) are shown.
    All blocks $S$ are explicitly shown. The multiplicity prefactors
    due to the different positions of $S$'s,
    c.f. eq. (\ref{eqq1}), are equal to 4 for (a), 4 for (b) and 1 for (c).
    The $A^j$'s are {\em not} included in $Q_1$, only the $G$'s and $S$'s. }
    
\label{fig6}
\end{figure}

\begin{figure}[tb]
  \includegraphics[width=9truecm]{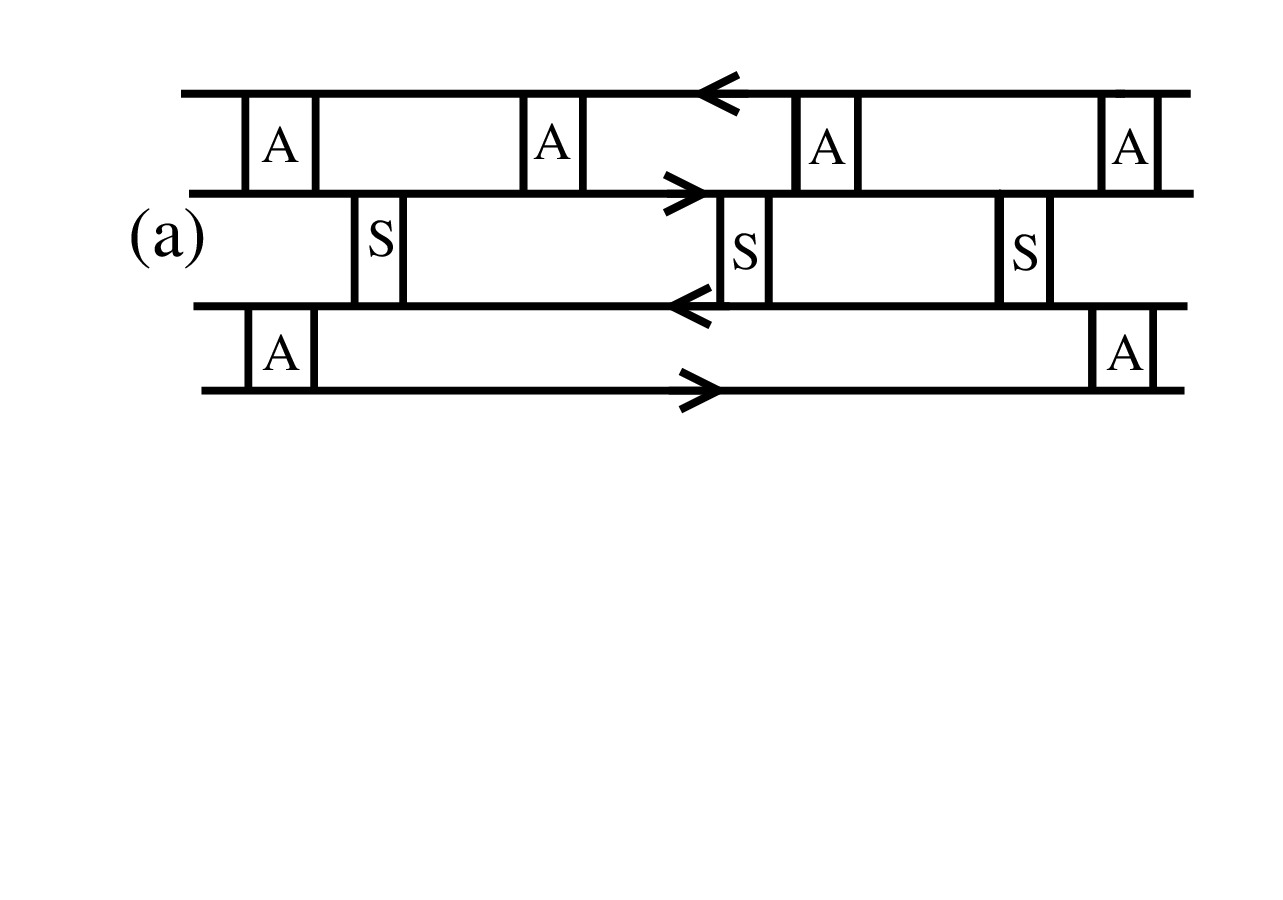}
  \includegraphics[width=9truecm]{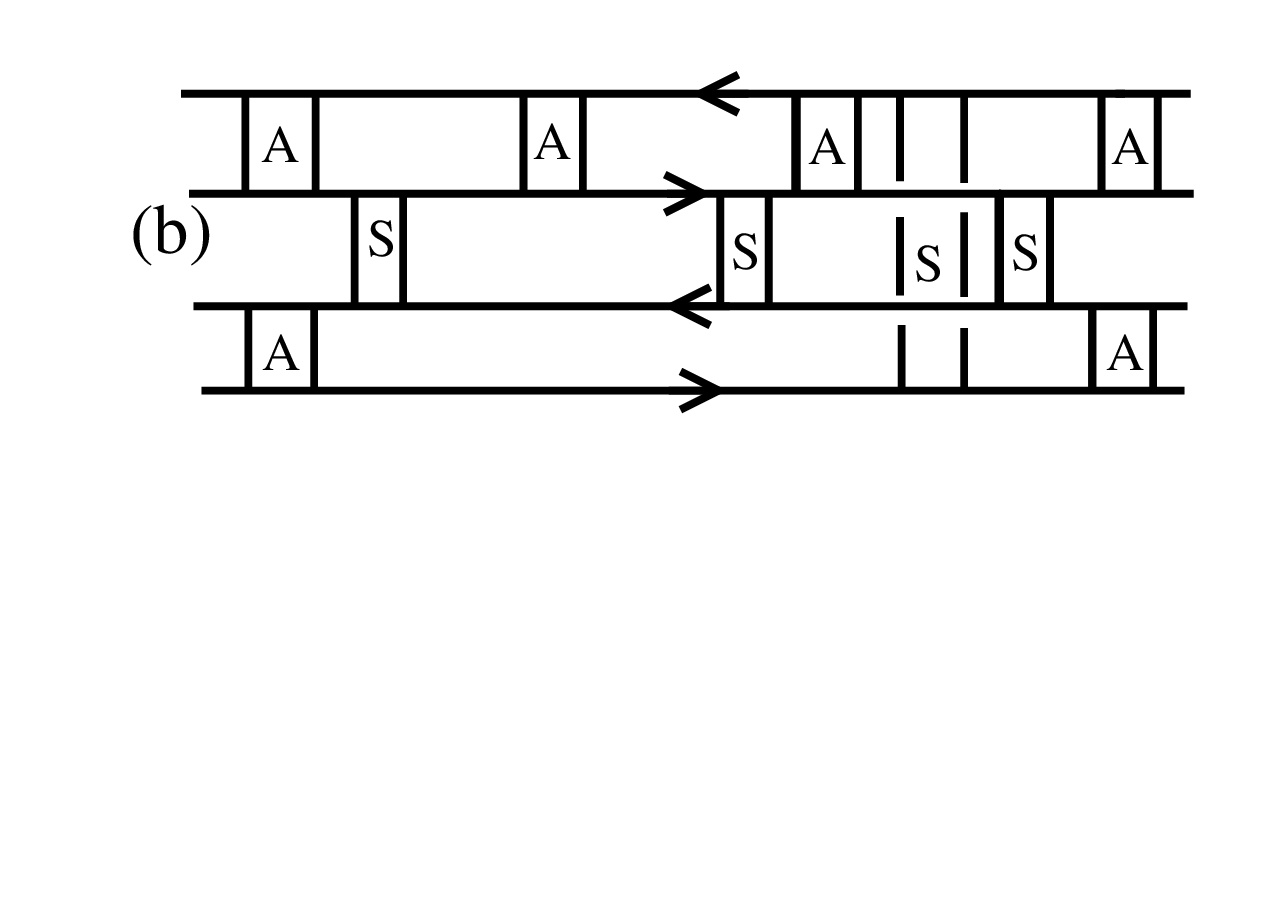}
  \includegraphics[width=9truecm]{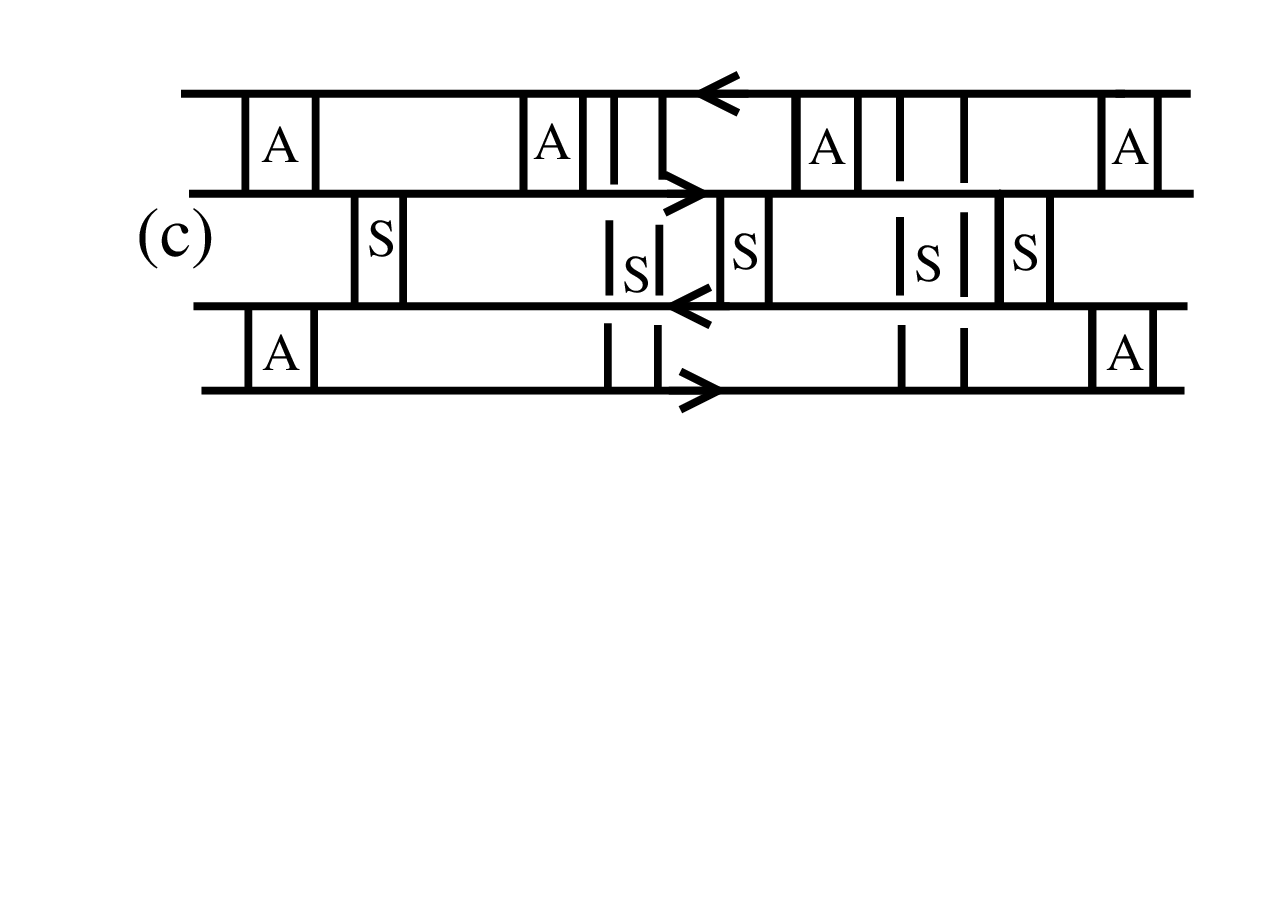}
  \includegraphics[width=9truecm]{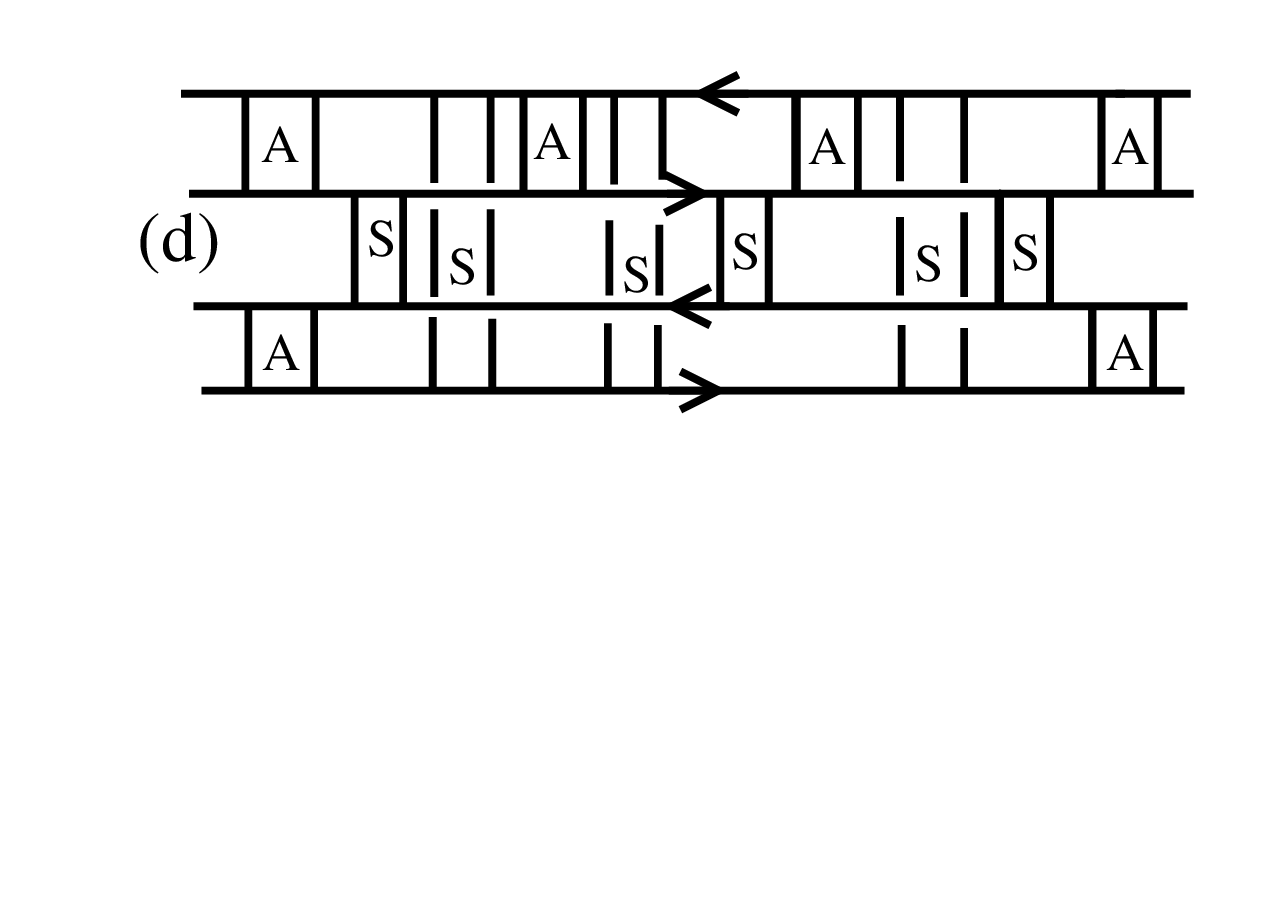}
  
  \caption{ Diagrams for the {\em internal} part $Q_2$, i.e. without the
    outer bubble lines. Four distinct contributions (a)-(d) are shown.
    All blocks $S$ are explicitly shown. The multiplicity prefactors
    due to the different positions of $S$'s, c.f. eq. (\ref{eqq2}), are
    equal to 8 for (a), 12 for (b), 6 for (c) and 1 for (d).
    The $A^j$'s are {\em not} included in $Q_2$, only the $G$'s and $S$'s. }
    
\label{fig7}
\end{figure}

The total contribution from the infinite series here is
\be
X_H = X_1(z) + X_2(z) \;\; ,  \label{exxh}
\ee
where 
$X_2(z) = -c_0  \; Q_2 \; f_2(z) $, with $(2m-3)$ terms of order $z^{2m-4}$,
and
\be
f_2(z) = \sum_{m=2}^{\infty} \frac{h_m \; z^{2 m}}{2m-1}
= -z^2 + z \tan^{-1}(z) \;\; .\;\;    \label{exf2}
\ee
Also $X_1(z)= -d_0  \; Q_1^2 \; f_1(z) $, with $(2m^2-7m+6)$ terms of 
order $z^{2m-5}$ and
\be
f_1(z) = \sum_{m=3}^{\infty} \frac{h_m \;(2m^2-7m+6) \; z^{2 m}}
{(2m-1)(2m-3)} = \frac{1}{2(1+z^2)}
\left\{ 3z^2+2z^4-3(z+z^3) \tan^{-1}(z) \right\}
\;\; ,\;\;    \label{exf1}
\ee
where $c_0=8  r_0 \; B_u^4 /P^4 \; l_0^2$ and $d_0=c_0/P$.
In the limit $z \rightarrow \infty$
these functions are $f_1(z) \simeq z^2$ and $f_2(z) \simeq -z^2$.

{\em Now we will examine closely some particular limits for $X_H$.}
First, for $\ep_F \tau \gg 1$ we may expand in
powers of $c_H$ ($c_H \propto \tau^{-4}  \propto s^4 $). Here we explicitly
consider that
$s^2>\sqrt{c_H}$. The specific case $s^2 \geq \sqrt{c_H}$ is treated below.
So we have
$S(k)=c_H \; \Gamma_k + O(c_H^2)$ and
\be
Q_1 =  r_1 \; c_H^2  + O(c_H^3) \;\; , \;\; 
Q_2 = r_2 \; c_H^3 + O(c_H^4) \;\; , \;\; 
P = P_0 \; c_H  +  O(c_H^2) \;\; . \;\;    \label{limet}
\ee
Here $r_1 = 4 \sum_{k<k_F} X_k^{-7} \; = (231/512) (N_F \; \pi/ s^{13})$,
$r_2 = 8 \sum_{k<k_F} X_k^{-10} \; = (12155/16384) (N_F \; \pi/ s^{19})$,
$P_0 = 2 \sum_{k<k_F} X_k^{-4} \; =(5/16) (N_F \; \pi/ s^7)$.

Using these in the limit $z \rightarrow \infty$ we have
\be
X_H \simeq (c_0 \; Q_2 - d_0 \; Q_1^2) \; z^2 = 
\frac{2 r_0 \; l_0^2 \; y_0 \; c_H}{P_0^3}  \;\; , \;\;   \label{xh1}
\ee
with $y_0 = P_0 \; r_2 -r_1^2 =\left\{(3707/20480) \simeq 0.1810 \right\}
(N_F^2 \pi^2/4 s^{26})$.

We note that in this limit $X_H$ in eq. (\ref{xh1}) is {\em independent}
of $B_u$. This is due to the combination $X_H \propto z^2 \; B_u^4$.
The same applies for $X_H$ in eq. (\ref{xh2}) below.

Next, we write $x_1 = s^2 -\sqrt{c_H}$ and $x_2 = s^2 +\sqrt{c_H}$,
and we consider the limit
\be
x_1 \rightarrow 0^+ \;\;,  \label{x1-0}
\ee
i.e. {\em $c_H$ is allowed to approach $s^4$ from below.}
As $c_H$ is a decreasing function of $H$ and $T$, this condition may only
apply for small $H$ and low $T$, i.e. in the region of the phase diagram
where the magnitude of the conductivity is {\em maximum}. 

Then, with $g=(\ef^2+s^2)^2-c_H$, we obtain the limiting forms
\be
P = \frac{p_{1A}}{x_1^{3/2} } + \frac{p_{1B}}{x_1 \; x_2 }
+ O(1/\sqrt{x_1})  \;\;,\;\;
\ee
with $p_{1A} = N_F \; \left(c_H^{1/2}-6 x_1\right) \;
\tan^{-1}\left( \ef/\sqrt{x_1} \right)/\left\{ 8 \; c_H^{1/2} \right\}$,
$p_{1B} = N_F \; \ef \; \left[c_H+s^2(s^2+\ef^2) \right] /\{4 g\}$,
\be
Q_1= \frac{q_{1A}}{x_1^{7/2} } + \frac{ q_{1B}}{x_1^3 \; x_2^3 } +
O(x_1^{-5/2})  \;\;,\;\;
\ee
with $q_{1A} = N_F \left\{ -512 \; s^6 + 1652 \; s^4 \; c_H^{1/2}
-1798  \;s^2 \; c_H +663 \; c_H^{3/2} \right\} \; 
\tan^{-1}\left( \ef/\sqrt{x_1} \right)/\left\{ 256 \;c_H \right\} $,
$q_{1B}=N_F \; \ef \left[ 6s^8 (91s^2+43 \ef^2) - 7c_H(196 s^6+93 s^4 \; \ef^2)
+c_H^2(946 s^2 + 453 \ef^2) \right]/\{ 384 \; g\}$, and
\be
Q_2 = \frac{q_{2A}}{x_1^6 \; x_2^5}+\frac{ q_{2B}}{x_1^5 \; x_2^{11/2}}
+\frac{q_{2C}}{x_1^5 \; x_2^5 }
+ O(x_1^{-4})  \;\;,\;\;
\ee
with $q_{2A}= N_F \left\{ -96096 \; s^{20} +23472 \; s^{18} \; c_H^{1/2}
+496464 \; s^{16} \; c_H -106684 \; s^{14} \; c_H^{3/2} -1033822 \; s^{12} \; c_H^2
+186043 \; s^{10} \; c_H^{5/2}  \right . $
$\left. + 1088757 \; s^{8} \; c_H^3 -149086 \; s^6 \; c_H^{7/2}
-584084 \; s^4 \; c_H^4 + 47263 \; s^2 \; c_H^{9/2} + 129789 \; c_H^5 \right\}
\tan^{-1}\left( \ef/\sqrt{x_1} \right)/\left\{ 16384 \; c_H^{3/2} \right\}$,
$q_{2B}= N_F \left\{ 96096 \; s^{20} +23472 \; s^{18} \; c_H^{1/2}
-496464 \; s^{16} \; c_H -106684 \; s^{14} \; c_H^{3/2} +1033822 \; s^{12} \; c_H^2
+186043 \; s^{10} \; c_H^{5/2}  \right . $
$\left. - 1088757 \; s^{8} \; c_H^3 -149086 \; s^6 \; c_H^{7/2}
+584084 \; s^4 \; c_H^4 + 47263 \; s^2 \; c_H^{9/2} - 129789 \; c_H^5 \right\}
\tan^{-1}\left( \ef/\sqrt{x_2} \right)/\left\{ 16384 \; c_H^{3/2} \right\}$, and
$q_{2C}= N_F \left\{ \ef \left(80637 c_H^5 -102009 \; s^4 \; c_H^4
-148193 \; s^8 \; c_H^3 + 387913 \; s^{12} \; c_H^2 -282716 \; s^{16} \; c_H
+70416 \; s^{20} \right)
+ \ef^3 \left( 141789 \; s^2 \; c_H^4  \right . \right .$
$\left. \left. -447258 \; s^6 \; c_H^3
+ 558129 \; s^{10} \; c_H^2 -320052 \; s^{14} \; c_H + 70416 \; s^{18}  \right)
\right\} /  \left\{24576 \; c_H \; g \right\}  $.

Now, taking $c_H \simeq s^4$ in $x_2$ and $g$, but {\em not} in $x_1$,
and $\ep_F \tau \gg 1$ yields $p_{1A}= N_F \pi/16$,
$q_{1A} = -15 N_F \pi \; s^2 /1536$ and $q_{2A} = 63 N_F \pi \; s^4 /32768 $.
Then the leading contribution in $1/x_1$ is
\be
X_H = \frac{2 r_0 \; l_0^2}{p_{1A}^2}  \;
\frac{ F_1}{x_1^{5/2}} \;\; , \;\;     \label{xh2}
F_1 = q_{2A} - q_{1A}^2/p_{1A} = \frac{77 \pi \; N_F}{32768} \;\;.
\ee

{\em Comparison between the series $W_H$ and $X_H$.}

The $z^{2m}$ term of the series $X_H$ dominates over the respective term
of the series $W_H$. The reason is the high
value of the numerical prefactor at any high enough power $z^{2m}$, which
{\em overcompensates} the missing factor $1/B_u^2$ from the two missing
$A^j$'s. For the limit $x_1 \rightarrow 0$ this is seen e.g. in the ratio
\be
\left(X_{2,m}/W_{1H,m}\right) = 8  B_u^2 \; q_{1a}^2 \; (2m-1)(2m^2-7m+6) /
\left\{ N_F^2 \; p_{1a}^4 \; l_0^2 \; (2m-3) \; x_1 \right\}  \;\; .
\ee
Here
the factor $1/x_1$ contributes as well. The ratio $\left(X_{1,m}/W_{1H,m}\right)$
behaves similarly.
For {\em finite} $x_1$ and $\ef \tau \gg 1$ - c.f. eq. (\ref{limet}) -
the ratio is
\be
\left(X_{2,m}/W_{1H,m}\right) = 8 r_1^2 \; B_u^2 \;(2m-1)(2m^2-7m+6) /
\left\{ P_0^4 \; l_0^2 \; (2m-3)  \right\}  \;\; ,
\ee
and similarly for the ratio $\left(X_{1,m}/W_{1H,m}\right)$.

However, the sign of the terms of the series $X_H$ {\em alternates} with $m$
as $h_m=(-1)^{m+1}$.
As a result, $X_H$ cannot dominate over $W_H$, and $W_H$ makes the main
contribution. Moreover, as shown above, for large $z$ the $X_H$ contribution
is limited in magnitude, as opposed to the $W_H$ contribution.
For completeness, we include both $W_H$ and $X_H$ in $A_H$ and $B_H$ below.

\begin{figure}[tb]
  \includegraphics[width=6truecm]{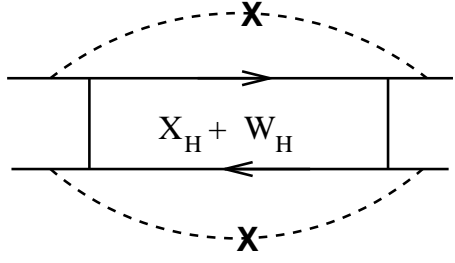}
  \caption{Diagrams for $A_H$ in eq. (\ref{qah}) and $B_H$ in eq. (\ref{qbh}).
 The dashed external lines with the cross stand for {\em impurity scattering}.
    $B_H$ corresponds to the case of one such scattering line, either above
    or below the central box, and also to the {\em absence} of these lines.
    $A_H$  corresponds to the presence of {\em both} scattering lines.}
    
\label{figab}
\end{figure}

\begin{figure}[b]
  \includegraphics[width=11truecm]{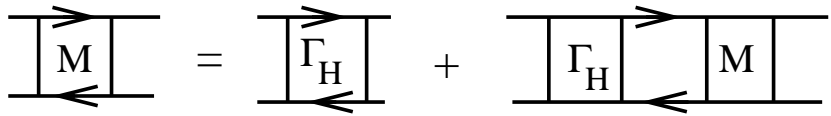}
  \caption{ Ladder diagrams for the block $M$ in terms of $\Gamma_H$.}
    
\label{figmg}
\end{figure}

\vspace{0.6cm}
{\bf 5. Calculation of the conductivity using the series $W_H$ and $X_H$}
\vspace{0.3cm}

With these series at hand, we further consider three types of diagrams
incorporating them. The ones with a single external impurity scattering
line and the ones {\em without} such a line - c.f. fig. \ref{figab} -
yield a {\em combined}
contribution $B_H \; G_R(k,\ep) G_A(k,\ep) $, where
\be
B_H = (W_H+X_H)  \; \left( 1+ \frac{\ef}{\pi s} \right) \;\; . \;\; \label{qbh}
\ee
The diagrams with two external impurity scattering lines - c.f.
fig. \ref{figab} - yield a contribution
$A_H$ with
\be
A_H = (W_H+X_H) / (4 \pi^2 \; s^2 )  \;\; .\;\;  \label{qah}
\ee

Hence in total we have
\be
\Gamma_H = A_H + B_H \; G_R(k,\ep) G_A(k,\ep) \;\; .
\ee

\begin{figure}[h]
  \includegraphics[width=6truecm,height=3truecm]{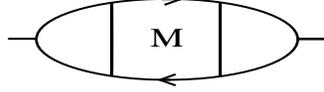}
  \caption{ Conductivity diagram for eq. (\ref{condm}).
    A block $M$ in the middle, and current vertices on each side.}
    
\label{figcm}
\end{figure}

Now we consider the total conductivity as
\be
\si = \si_D + \si_A(H) \;\; , \;\; \label{sit0}
\ee
where, $\si_D$ is the Drude term, and for $T \rightarrow 0$, $\si_A(H)$
is given by
\be
\si_A(H) = \frac{ e^2}{3 \pi \; m^2} \int_0^{k_F} \; d{\vec k} \; k^2 
\left \{ G_R(k,0) G_A(k,0) \right \}^2 \Gamma_{H}  \;\; . \;\; 
\ee
Calculating the integral we obtain
\be
\si_A(H) = \frac{e^2 \; N_F}{3\pi \; m} \left\{
\frac{\ef}{8s^5} \left( 4\; A_{H} \; s^2
+ 3\; B_{H} \right) \text{tan}^{-1}\left(\frac{\ef}{s}\right)
-\frac{B_{N}}{4 s^4} + \frac{5 \; B_{H}}{8 \; s^2 (s^2+\ef^2)} \right\}
\;\; .\;\;           
\ee
Considering the term $W_{H}$ {\em only} in the low disorder limit
$\ep_F \tau \gg 1$, we notice that $\si_A(H)$ is
\be
\si_A(H) = s_0 \; f(z) \;\;    \label{sbas}
\ee
with, c.f. eqs. (\ref{w1h}),
\be
f(z) = z \; I_1(z) \;\; , \;\;
\ee
and
\be
s_0 = \frac{e^2 \; N_F \; \ef^2 \; r_0 \; l_0^2}{16 \; m \; s^6}  \;\; . \;\;
\ee

Here $l_0 = r \; m_u \; \lambda$, $\lambda^4=1+1/\Lambda_u^4$, 
$r=4/(3 \; \tau_S)=b_1 \; s$ with $b_1=O(1)$, $M_{u0}=r\; m_u$,
$m_u=U u $, $C_u = r \; c_u$.

The Drude term is
\be
\si_D = \frac{ e^2}{3 \pi \; m^2} \int_0^{k_F} \;
d{\vec k} \; k^2 \; G_R(k,0) G_A(k,0) \simeq
\frac{e^2 \; N_F}{3\pi \; m}
\frac{\ef}{s}\text{tan}^{-1} \left(\frac{\ef}{s} \right)
\;\;.\;\;    \label{sidr}
\ee

It turns out that eq. (\ref{sbas}) has very interesting properties, relevant
to the XMR experiments. For intermediate magnetic fields $H$ it yields
XMR proportional to $H^b$, $b \geq 1$. For higher $H$ the behavior changes
with $b \rightarrow 2$ as $H$ grows. This is how the variety of the values of
the exponent $b$ observed experimentally can be explained.

Then we consider the infinite ladder sum of $\Gamma_H$ - c.f. fig. \ref{figmg}
\be
M(k,\ep) = \frac{ \Gamma_H} { 1- \Gamma_H \; G_R(k,\ep) G_A(k,\ep)} \;\;. \;\;
\label{eqm}
\ee

For $T \rightarrow 0$ the total conductivity due to this mechanism is given by
\be
\si_M = \frac{ e^2}{3 \pi \; m^2} \int_0^{k_F} \; d{\vec k} \; k^2 
\left \{ G_R(k,0) G_A(k,0) \right \}^2 M(k,0) = -\si_D+\si_C(H) \;\; .\;\;
\label{condm}
\ee
Eq. (\ref{condm}) is a Baym-Kadanoff conserving approximation 
\cite{baym},\cite{gmr},\cite{mit}. This is seen by considering the free
energy diagrams which result by removing the two current vertices from
the conductivity diagrams above. It turns out then that the only
way to reintroduce the current vertices and obtain an {\em enhanced} 
conductivity yields precisely the original diagrams. 
$(-\si_D)$ is {\em minus} the 
well known Drude term.
This results from $\Gamma_H$ in the denominator of $M(k,\ep)$, and its
origin is more clearly seen by considering the limit
$\Gamma_H \rightarrow \infty$ in eq. (\ref{condm}). As a result, the Drude 
term {\em drops out} of the total conductivity $\si(H)$, 
which is given by the highly $H$-sensitive term $\si_C(H)$. 
In turn, this is how giant magnetoresistance appears
as a {\em bulk effect} in these systems, as $\si_C(H)$ decreases quickly with
increasing $H$. This is yet another impressive manifestation of quantum 
interference effects at the macroscopic scale. The cooperon incorporates
the effects of counter-propagating electron paths \cite{lee}, which interfere 
destructively with the increasing magnetic flux enclosed (due to $H$).
The present mechanism acts as a 'lense', which drastically magnifies the
interference effects.

We have
\be
\si(H) = \si_D+\si_M  = \si_C(H)   \;\; , \;\; 
\ee
\be
\si_C(H) = \frac{ 2 e^2 \; N_F}{3 \pi m \; S_H}
\left \{ \frac{\ef Y_+}{R_+} \tan^{-1} \left(\frac{\ef}{R_+} \right)
- \frac{\ef Y_-}{R_-} \tan^{-1} \left(\frac{\ef}{R_-} \right)
+Y_+ \; \ln\left(\frac{ R_+^2}{R_+^2 + \ef^2} \right)    
-Y_- \; \ln\left(\frac{ R_-^2}{R_-^2 + \ef^2} \right)  \right\} \;\; , \;\;
\label{sigm}
\ee
\be
R_{\pm} = \sqrt { s^2 - Y_{\pm} }  \;\; , \;\;
Y_{\pm}= (A_H\pm S_H)/2 \;\; , \;\; S_H= \sqrt{ A_H^2+4 B_H } \;\;. \label{exry}
\ee
Here we assumed a constant $N_F$, and
the $ln$ terms are much smaller than the first terms and may be ommitted.

This relation yields GMR as a function of $H$ with $\rho(H)=1/\si(H)$.
The magnitude of GMR increases with the weakening of disorder, i.e. for
increasing $\ef \tau$. From eq. (\ref{sigm}), GMR can be attributed to
$R_+(H=0) \rightarrow 0^+$, i.e. $Y_+(H=0) \rightarrow s^2$, so that a
substantial contribution comes from $1/R_+ $. As $H$ increases, $c_H$ is
reduced, and so are the series $X_H$ and $W_H$, yielding a lower $\si(H>0)$.

We defer a further discussion on GMR to the end of section 6, after
eq. (\ref{exww}).

\begin{figure}[h]
  \includegraphics[width=8truecm]{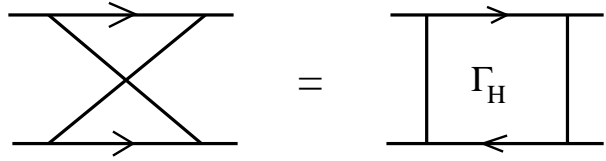}
  \caption{ The 'bow-tie' shaped $\Gamma_H$. It is derived from $\Gamma_H$
    just by rotating around the lower line. Notice the sense of the arrows
    shown.  }

\label{figbow}
\end{figure}

\vspace{0.6cm}
{\bf 6. More complicated diagram series and the respective conductivity
}
\vspace{0.3cm}

In this section we introduce additional diagrammatic contributions, which
include the block $\Gamma_H$ in {\em more complicated} series than
in the previous section.

First, we introduce the 'bow-tie' shaped $\Gamma_H$, shown in fig. \ref{figbow}.
It is derived from $\Gamma_H$ just by rotating around the lower line. Then,
we form a new effective interaction by combining two or more such bow-tie
$\Gamma_H$'s in series, as shown in fig. \ref{figrk}. We draw attention to
the precise direction of arrows in the figure.
We note that the term
with only one $\Gamma_H$ is already contained in the series for $M$
as in eq. (\ref{eqm}). Explicitly we have (with $\Gamma_k$ defined
right before eq. (\ref{eqs}) )
\be
R(k) = \Gamma_H^2 \;  \Gamma_k \; \sum_{m=0}^{\infty} \;
\left( \Gamma_H \;  \Gamma_k \right)^m = \frac{ \Gamma_H^2 \;  \Gamma_k }
{1- \Gamma_H \;  \Gamma_k }   \;\;.  \label{eqrk}
\ee

Subsequently we will form laddder diagrams which include both $R(k)$'s
and $M$'s. These are split into two discrete contributions $L_1$ and $L_2$
obeying the coupled Bethe-Salpeter equations in fig. \ref{fil12}
\bea
L_1 = \Gamma_H + \Gamma_H \; \Gamma_k \; L_1 + R(k) \; \Gamma_k \; L_2
 + L_2\; \Gamma_k \;  R(k) \;\;,
\label{exl1} \\
L_2 = R(k) + R(k) \; \Gamma_k \; L_1 + L_1  \; \Gamma_k \; R(k)
+ \Gamma_H \; \Gamma_k \; L_2 +  L_2 \; \Gamma_k \; \Gamma_H  \;\; .
\label{exl2}
\eea
We note that without the terms $R \; \Gamma_k \; L_2 $ eq. (\ref{exl1})
yields precisely $L_1 = M(k,\ep)$.

\begin{figure}[t]
  \includegraphics[width=6truecm]{xfig-rser.eps}
  \caption{ Diagrams for the block $R(k)$ in eq. (\ref{eqrk}). We sum up
    the infinite series which
    starts with two 'bow-tie' shaped $\Gamma_H$'s - c.f. fig. \ref{figbow}.
    The term with only one $\Gamma_H$ is already contained in the series
    for $M$ as in eq. (\ref{eqm}). Notice the sense of the arrows.}
    
\label{figrk}
\end{figure}

\begin{figure}[h]
  \includegraphics[width=10truecm]{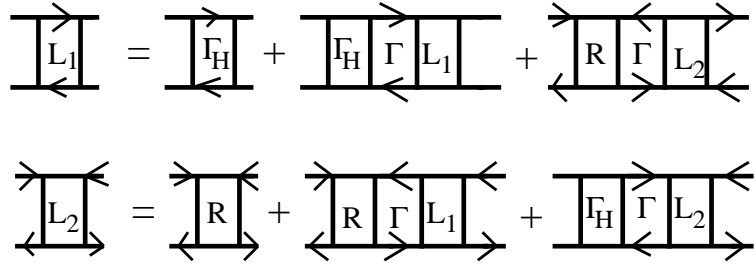}
  \caption{ Coupled equations for the propagators $L_1$ and $L_2$.
    We do {\em not} show the full, i.e. mirror-symmetric, terms, omitting
$L_2 \; \Gamma_k \; R $, $L_1 \; \Gamma_k \; R$, and 
$L_2 \; \Gamma_k \; \Gamma_H $. Notice the {\em precise sense} of the arrows.  }
    
\label{fil12}
\end{figure}

\begin{figure}[b]
  \includegraphics[width=8truecm,height=3.3truecm]{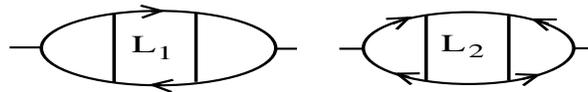}
  \caption{ Conductivity diagrams for the two propagators $L_1$ and $L_2$. 
  }
    
\label{ficol}
\end{figure}

The contributions to the conductivity from $L_1$ and $L_2$ are shown in
fig. \ref{ficol}. Taking
\be
L(k) = L_1 + L_2 \;\;,
\ee
these two diagrams can be combined into a single expression.
For $T \rightarrow 0$ the total contribution from $L(k)$ is given by
\be
\si_L = \frac{ e^2}{3 \pi \; m^2} \int_0^{k_F} \; d{\vec k} \; k^2 
\left \{ G_R(k,0) G_A(k,0) \right \}^2 L(k)    \;\;.
\label{conll}
\ee

We also subtract relevant contributions to avoid double counting.
Again this is a Baym-Kadanoff conserving approximation, like the
equivalent eq. (\ref{condm}).
Calculating the integral, with the assumption of a constant $N_F$, we obtain
\be
\si_L = -\si_D + \si_*(H) \;\; ,\;\;          \label{sigm1} 
\ee
where
\be
\si_*(H) = \frac{  e^2 \; N_F \; \ef \; A_*}{3 \pi m \; T_H }
\left \{ \frac{ K_+}{V_+} \tan^{-1} \left(\frac{\sqrt{2} \; \ef}{V_+} \right)
- \frac{ K_-}{V_-} \tan^{-1} \left(\frac{\sqrt{2} \; \ef}{V_-} \right)
 \right\} \;\; . \;\;
\label{sigml}
\ee
Here $A_* = (c_*^2-3+c_*-1/c_*)/\{ \sqrt{2} \; (c_*-d_*) 
\; [(a_*-c_*)^2+b_*^2] \}=0.401872$. 
The constants arise as solutions of a quartic equation, and are given by 
$c_*=2.748403$, $a_*=1.234465$, $b_*=1.350696$, $d_*=-0.2173337$. 
Also $V_{\pm} = \sqrt { 2 s^2 - \sqrt{c_*} \; K_{\pm} }$, 
$K_{\pm}= \sqrt{c_*} \; A_H \pm  T_H$, $T_H= \sqrt{ c_* \; A_H^2+4 B_H }$.
There are also $ln$ terms, which are much smaller than the terms shown,
and may be ommitted.

GMR is obtained when $\si_*(H \rightarrow 0)$ is significantly enhanced.
This is the case especially for
\be
V_+^2 = 2 s^2 - \sqrt{c_*} \; K_+ \rightarrow 0 \;\; .
\ee
Using eqs. (\ref{qah}) and (\ref{qbh}), this happens for
\be
W_H + X_H = \frac{s^4}{c_*} \; \frac{1}{\ef/(\pi \; s) + 1 +1/(4 \pi^2)} = w_*
\;\; . \label{exww}
\ee
We note here that the equivalent relation $R_+^2  \rightarrow 0$ in the
previous section yields $W_H + X_H = c_* \; w_*$. I.e. the consideration
of all these additional diagrams in the present section {\em reduced} the
appropriate value of $W_H + X_H$, required for GMR, by a factor of $c_*$.

As for eq. (\ref{sigm}), $\si(H)$ is a decreasing function of $H$ due to
the decay of $c_H$ with $H$. This yields {\em positive} MR.
Further, we see that the cleaner the sample, i.e. the higher the ratio
$\ef/s$ is, the smaller $w_*$ is. Also $w_* \propto s^4$. Together these
facts imply that {\em the magnitude of GMR is higher in cleaner samples,
in agreement with experiments} \cite{cavsb}. This also implies that in
cleaner samples the conditions, due to small values of $B_u$ and otherwise
as elaborated upon in eqs. (\ref{xh1}) and (\ref{xh2}) in Section 4, are
easier to satisfy, making GMR more likely to appear therein.

The discussion about the $H^b$ dependence of XMR, given after eq. (\ref{sidr}),
applies here as well.
The anisotropy, seen in the experiments, with the direction of the $H$ 
is easily explained in terms of the anisotropic diffusion coefficient
entering in $c_H$. The main $T$ dependence comes from the fast rise
of $\tau_{\phi}^{-1}(T)$ with $T$, which yields the respective decay of GMR. 

\vspace{0.5cm}
PtBi$_2$ deserves a special mention. Pt was shown experimentally in \cite{ptf}
to display GMR, in accordance with the value of $u$ for (fcc) Pt following our
theory \cite{gmr}. Interestingly, PtBi$_2$ in ref. \cite{ptbi}
has the crystal structure of (fcc) Pt with an additional Bi matrix. The 
samples in \cite{ptbi} display possibly the biggest XMR found, as mentioned
in the first paragraph of the Introduction.

Moreover, we draw attention to the fact that XMR and GMR have appeared 
in a number of compounds which contain Pt and/or Sc, both of which are
expected to display these effects \cite{gmr}. These are PtBi$_2$ \cite{ptbi}
and PtSn$_4$ \cite{ptsn} for XMR, and ScPtBi \cite{scpt} and
PtBi$_2$ \cite{ptbi-b} for GMR. Together with the data from Rh \cite{ptf},
this set of data supports our theoretical model.

\vspace{0.3cm}

\begin{figure}[t]
  \includegraphics[width=9truecm]{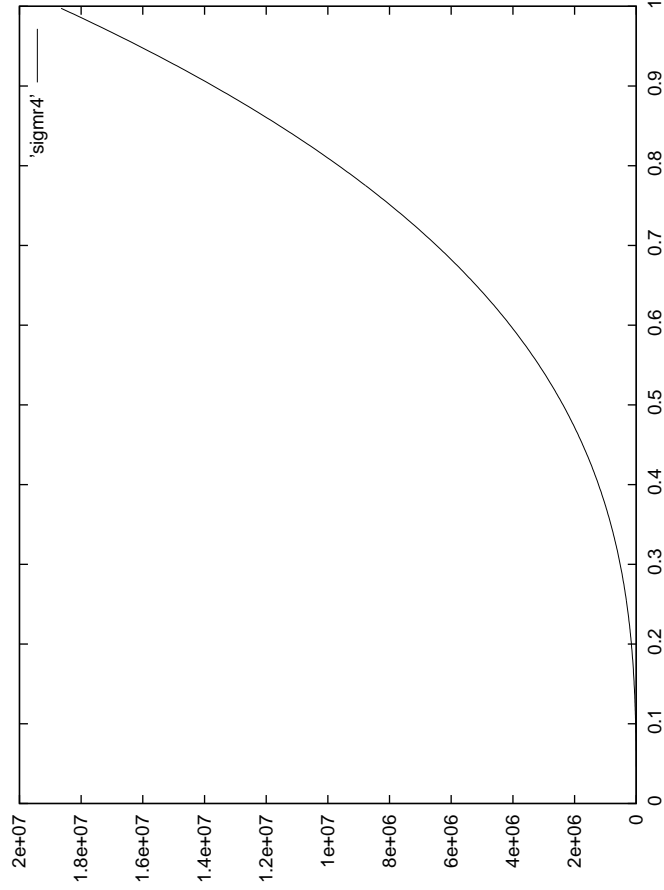}
  \caption{ Plot of XMR$=\rho(H)/\rho(0)-1=\si(0)/\si(H)-1$
    from eqs. (\ref{sigml}),(\ref{sigm1}) as a
    function of the magnetic field $H$. 
    In this figure XMR {\em exceeds} 1.8 $10^7$. For large enough
    $H$ XMR saturates
    due to the decay of $c_H$ in eq. (\ref{lch}). For the same reason
    XMR vanishes at elevated $T$.}
    
\label{fxmr}
\end{figure}

We plot the conductivity from our eqs. (\ref{sigm1}),(\ref{sigml}) in
fig. \ref{fxmr}. $\si_*(H)$ can be much bigger than the Drude term $\si_D$.
The resistivity eventually saturates
at some high value of the field $H$. $\si_*$ decreases with $H$ until it
becomes
smaller than $\si_D$, at which point saturation is reached. For similar reasons
the resistivity should actually saturate at high enough $H$ in all cases.

\vspace{0.3cm}

In order to consider $z$ values bigger than 1, which arise, say,
from $B_u \rightarrow 0$, we need a reasonable estimator for $W_H$, i.e.
$I(z)$ (and likewise for $X_H$).
$I(z)$ was calculated to infinite order from Feynman diagrams, and,
physically speaking, should {\em converge} for $z \gg 1$. One possible
way to estimate $I(z>1)$ is via the use of Pad\'e approximants \cite{bak}.
However, they introduce both unphysical poles and an asymptotic behavior for
$z \gg 1$ which may be unphysical. A recent better approximation method
appears to be the use of hypergeometric functions \cite{mera,sand}. They do
not have poles at finite $z$, their asymptotic behavior for $z \gg 1$ is
reasonable, and they dispose a flexible and general character, as several
functions are special cases of them.
Specifically, we approximate
\be
I_i(z) \simeq y_i
   {\text Re} \left\{ _2 F_1(a_i,b_i,c_i, d_i \; z^2) \right\} \;\;,
\ee
for both $i=1,2$ by the hypergeometric function
\be
_2 F_1(a,b,c, d \; x )= \sum_{n=0}^{\infty} \; \frac{(a)_n \; (b)_n}{n! \; (c)_n}
\; d^n \; x^{n}  \;\;, \;\;
\ee
where $(a)_n=a(a+1)...(a+n-1)$, $(a)_0=1$ and Re denotes the real part,
as the conductivity we are interested in here is manifestly real
(and $_2 F_1(a,b,c, d \; x )$ for $d \; x > 1$ is complex in general,
and given that the coefficients $a_i$ and $b_i$ turn out to be complex).
The coefficients $y_i$ adjust the overall scale as the first term of $_2 F_1$
is 1. They are $y_1=4$ and $y_2=4/9$.
We obtain the coefficients $a_i, b_i, c_i, d_i$ by equating the
coefficients of $y_i  \; _2F_1$ and $I_i$ for $n=1-4$. Thus we have 
$ a_1=\left(2019-i\sqrt{404039}\right)/4600, \;
 b_1=\left(2019+i\sqrt{404039}\right)/4600, \;
 c_1=1461/764, \; d_1=575/573$ and
 $a_2=\left(11397 - 9 i \sqrt{67191}\right)/10976, \;
 b_2=\left(11397 + 9 i \sqrt{67191} \right)/10976, \; c_2=17125/5464, \;
 d_2=2058/683$. We note that $a_i$ and $b_i$ are complex congugate. 

 Plotting $z I(z)$ we see that the term $z^2 \; I_2(z)/\Lambda_u^4$ is
 much smaller than $I_1(z)$ and may be ignored. We see that
 max $z I_1(z)\simeq 2.384$ for $z=13$.

\vspace{.5cm}
{\bf 7. Overview}
\vspace{0.3cm}

In this work we developed further our original diagrammatic model \cite{gmr}
for GMR, aiming to explain the XMR experiments already cited. Our model
explains the salient features of XMR. Namely, the big magnitude of
the effect as a function of the magnetic field $H$, its gradual
decay with increasing temperature T, and its anisotropy as a function
of the direction of $H$. The magnitude of XMR is higher in cleaner, i.e.
less disordered, samples. As stated after eq. (\ref{sidr}), for moderate $H$
our mechanism may yield XMR linear in $H$, which for higher $H$ may switch
to $H^b$, $b \sim 2$ etc., in agreement with experiments. Overall,
our mechanism acts in the bulk, {\em not} in the surface of the samples, 
and relies on a big conductivity for $H=0$, which is quickly reduced 
with increasing $H$.
All these are in good agreement with experiments. In section 6
we discussed the special role of Pt, Sc and Rh containing compounds.

A physical interpretation of XMR as a
manifestation of quantum interference effects at the macroscopic scale is
presented in section 5, after eq. (\ref{condm}).

\vspace{.3cm}
$^*$ e-mail : kast@iesl.forth.gr ; giwkast@gmail.com


\begin{thebibliography}{99}
\bibitem{gm1}
We give the first 16 $g_n$'s :
$g_1=-1,g_2=1,g_3=1,g_4=-1,g_5=-1,g_6=1,g_7=1,g_8=-1,
g_9=-1,g_{10}=1, \\ g_{11}=1,g_{12}=-1,g_{13}=-1,g_{14}=1,g_{15}=1,g_{16}=-1$, etc.
$g_n$'s are periodic in $n$ with a period $\Delta n=4$.
\end{thebibliography}
\end{document}